\documentclass[12pt]{article}
\usepackage{amsfonts}
\usepackage{amsmath}
\usepackage{amsthm}

\advance \topmargin by -\headheight
\advance \topmargin by -\headsep
     
\evensidemargin \oddsidemargin
\def\br{\begin{eqnarray}}
\def\er{\end{eqnarray}}
\def\be{\begin{equation}}
\def\ee{\end{equation}}

\def\({\left(}
\def\){\right)}

\relax

\def\a{\alpha}

\def\g{\gamma}

\def\h{ \frac{1}{2}  }

\def\pa{\partial}

\def\lie{{\cal G}}

\def\NPB#1#2#3{{\sl Nucl. Phys.} {\bf B#1} (#2) #3}

\def\PLA#1#2#3{{\sl Phys. Lett.} {\bf #1A} (#2) #3}
\def\PLB#1#2#3{{\sl Phys. Lett.} {\bf #1B} (#2) #3}

\begin{document}

\begin{center}
{\large\bf  A Class of Soliton  Solutions for the ${\mathbf N= 2}$  
Super}\\ {\large\bf mKdV/Sinh-Gordon Hierarchy }
\end{center}
\normalsize
\vskip .4in

\begin{center}
 H. Aratyn

\par \vskip .1in \noindent
Department of Physics \\
University of Illinois at Chicago\\
845 W. Taylor St.\\
Chicago, Illinois 60607-7059\\
\par \vskip .3in

J.F. Gomes,  L.H. Ymai and A.H. Zimerman

\par \vskip .1in \noindent
Instituto de F\'{\i}sica Te\'{o}rica-UNESP\\
Rua Pamplona 145\\
01405-900 S\~{a}o Paulo, Brazil
\par \vskip .3in

\end{center}

\begin{abstract}
Employing the Hirota's method, a class of soliton solutions for the $N=2$
super mKdV equations is proposed in terms of a single Grassmann
parameter.  Such solutions are shown to satisfy two copies of $N=1$
supersymmetric mKdV equations connected by nontrivial algebraic identities.
Using the super Miura transformation, we obtain solutions of the $N=2$ super
KdV equations.  These are shown to generalize solutions 
derived previously.  By using the mKdV/sinh-Gordon hierarchy properties we generate
the solutions of the $N=2$ super sinh-Gordon as well.
\end{abstract}

\vskip 1cm

The  supersymmetric $N=2$ Sinh-Gordon model was  first introduced in \cite{ik} and \cite{ku}. Moreover,  
in   \cite{ik} the supersymmetric $N=2$ mKdV and its Miura 
transformation to the supersymmetric $N=2$ KdV was also discussed.

In an algebraic approach, integrable hierarchies are 
defined by decomposition of an affine Lie algebra $\lie $ 
into  graded subspaces by a grading operator $Q$ and 
further specified by  a constant grade one element $E$.  
Such graded structure provide  a systematic way to obtain
solutions of the zero curvature equation  for a 
corresponding Lax operator. 
For each grade one finds a solution, which corresponds to 
a different time evolution  $t= t_k$ and hence to a 
different  nonlinear evolution equation.  
In particular, 
supersymmetric integrable hierarchies  require  
the decomposition of a twisted affine superalgebra.
In refs. \cite{nosso1} and \cite{nosso2},  the  half integer  
decomposition of affine $\hat{sl}(2,2)$ was discussed and  
the equations of motion for the $N=2$ super 
sinh-Gordon and mKdV were derived and shown to correspond to 
different  time evolutions of the same hierarchy.
The  algebraic structure behind the hierarchy ensures  
universality among the solutions of different equations of motion.
In fact, apart from changes of field variables, the space-time 
dependence of the $(2n+1)$-th member of 
the mKdV/sinh-Gordon hierarchy is given by
\br \rho ^{\pm}(x, t_{2n+1}) = \exp \( \pm (2\g x +
2\g^{2n+1}t_{2n+1})\)
\label{i1} 
\er
and the soliton solutions of different equations of motion  
within the same hierarchy differ only by its space-time form 
specified by (\ref{i1}) while maintaining similar functional form.
 
In \cite{gs} a class of soliton solutions for the 
supersymmetric $N=2$ KdV with one Grassmannian parameter was 
obtained employing Hirota's method.  
In this paper, we extend the construction of soliton solutions to
the supersymmetric $N=2$ mKdV model.  
The advantage of the method is that it also yields solutions 
to $N=2$ super sinh-Gordon model as the space-time dependence of 
solutions is provided by universality of solutions ensured by 
the fact that both models are embedded within the same hierarchy. 

By employing super Miura transformation  we arrive at a more general 
class of $N=2$ super KdV equation which for a particular choice 
of parameters agrees with the one obtained in \cite{gs}.

The $N=2$ super mKdV model is described by 
the $t=t_3$ flow of the  affine $\hat{sl}(2,2)$ hierarchy 
(see \cite{nosso1}) :
\be
\begin{split}
4\partial_{t_3}\psi_1&=\partial_x^3\psi_1-3(u_1^2+u_3^2)\partial_x\psi_1-\frac{3}{2}\partial_x(u_1^2+u_3^2)\psi_1-3\partial_x(u_1u_3)\psi_3\\
4\partial_{t_3} u_1&=\partial_x^3 u_1 +
\partial_x\left[-2u_1^3+3u_1(\psi_1\partial_x\psi_1-\psi_3\partial_x\psi_3)-3u_3\partial_x(\psi_1\psi_3)\right]\\
4\partial_{t_3}\psi_3&=\partial_x^3\psi_3-3(u_1^2+u_3^2)\partial_x\psi_3-\frac{3}{2}\partial_x(u_1^2+u_3^2)\psi_3-3\partial_x(u_1u_3)\psi_1\\
4\partial_{t_3} u_3&=\partial_x^3 u_3 +
\partial_x\left[-2u_3^3-3u_3(\psi_1\partial_x\psi_1-\psi_3\partial_x\psi_3)+3u_1\partial_x(\psi_1\psi_3)\right]
\, .
\label{1} 
\end{split}
\ee
The $N=2$ super sinh-Gordon model belongs to the same hierarchy
but with the flow parameter $t = t_{-1}$ 
for which one finds the following evolution equations :
\br
\partial_{t_{-1}}\partial_{x}(\phi_{1}\pm \phi_3)&=&4\sinh(\phi_{1}\pm \phi_3)\cosh(\phi_{1}\mp \phi_3)-4(\psi_{1}\pm \psi_3)(\bar{\psi}_{1}\pm \bar \psi_{3})\sinh(\phi_{1}\mp \phi_3),\nonumber\\
\partial_{t_{-1}}(\psi_{1}\pm \psi_3)&=&-2(\bar{\psi}_{1}\mp \bar
\psi_3)\cosh(\phi_{1}\pm \phi_3)\, ,
\label{2} 
\er
where $\bar \psi_{1,3}$ are  auxiliary fields satisfying
\br
\partial_{x}(\bar{\psi}_{1}\pm \bar \psi_3)&=&-2(\psi_{1}\mp
\psi_3)\cosh(\phi_{1}\pm \phi_3)\, .
\label{4}
\er
The fact that both integrable models belong to the same hierarchy is expressed by 
relation 
\br
u_i = -\pa_x \phi_i, \quad i = 1,3\, .
\label{5}
\er
Define now the superfields
\br
\chi_1=\psi_1+\theta u_1,\quad \quad 
\chi_3=\psi_3-\theta u_3,
\label{6}
\er
and the superderivative
\br
D=\partial_{\theta}+\theta\partial_x, \qquad D^2=\partial_x,
\label{7}
\er
The $N=2$ supersymmetric mKdV equations can be recast as
\be \begin{split}
4\partial_{t_3}\chi_1&=\partial_x^3\chi_1+D\left[-2(D\chi_1)^3+3(\chi_1\partial_x\chi_1-\chi_3\partial_x\chi_3)D\chi_1+3D\chi_3\partial_x(\chi_1\chi_3)\right]
\\
4\partial_{t_3}\chi_3&=\partial_x^3\chi_3+D\left[-2(D\chi_3)^3-3(\chi_1\partial_x\chi_1-\chi_3\partial_x\chi_3)D\chi_3-3D\chi_1\partial_x(\chi_1\chi_3)\right]
\label{8}
\end{split}
\ee
%which can be rewritten as
%\br
%4\partial_{t_3}\chi_1&=&\partial_x^3\chi_1-3\left[(D\chi_1)^2+(D\chi_3)^2\right]\partial_x\chi_1+3\chi_3\partial_x(D\chi_1D\chi_3)\nonumber\\
%&&-3\chi_1(D\chi_1D\partial_x\chi_1+D\chi_3D\partial_x\chi_3)\nonumber \\                                                                      4\partial_{t_3}\chi_3&=&\partial_x^3\chi_3-3\left[(D\chi_1)^2+(D\chi_3)^2\right]\partial_x\chi_3+3\chi_1\partial_x(D\chi_1D\chi_3)\nonumber\\
%&&-3\chi_3(D\chi_1D\partial_x\chi_1+D\chi_3D\partial_x\chi_3)
%\label{9}
%\end{eqnarray}

We now introduce the following tau functions :
\br
\chi_1=D\mathrm{ln}\left(\frac{\tau_1}{\tau_2}\right), \qquad
\chi_3=D\mathrm{ln}\left(\frac{\tau_3}{\tau_4}\right)
\label{10}
\er
and the Hirota's derivatives
\be
\begin{split}
{\bf SD}_{t_3}(\tau_1.\tau_1)&=2(D\partial_{t_3}\tau_1\,\tau_1-D\tau_1\partial_{t_3}\tau_1),\\
{\bf SD}_x(\tau_1.\tau_1)&=2(D\partial_x\tau_1\,\tau_1-D\tau_1\partial_x\tau_1),\\
{\bf SD}_x^3(\tau_1.\tau_1)&=2(D\partial_x^3\tau_1\,\tau_1-3D\partial_x^2\tau_1\partial_x\tau_1
+3D\partial_x\tau_1\partial_x^2\tau_1-D\tau_1\partial_x^3\tau_1),\\
{\bf D}_x^2(\tau_1.\tau_2)&=\partial_x^2\tau_1\,\tau_2-2\partial_x\tau_1\partial_x\tau_2+\tau_1\partial_x^2\tau_2,\\
{\bf D}_x^2(\tau_1.\tau_1)&=2\left[\partial_x^2\tau_1\,\tau_1-(\partial_x\tau_1)^2\right],\\
\bar{{\bf D}}(\tau_a.\tau_b)&=D\tau_a\partial_x\tau_b-D\tau_b\partial_x\tau_a,
\;\; a \neq b =1,2,3,4 \, .
\label{11}
\end{split}
\ee
The first of equations in (\ref{8}) becomes
\be
\begin{split}
&2\left[\frac{{\bf SD}_{t_3}(\tau_1.\tau_1)}{\tau_1^2}
-\frac{{\bf SD}_{t_3}(\tau_2.\tau_2)}{\tau_2^2}\right]
=\frac{{\bf SD}_x^3(\tau_1.\tau_1)}{2\tau_1^2}-\frac{{\bf SD}_x^3(\tau_2.\tau_2)}{2\tau_2^2}\\
&-\frac{3}{2}\left[\frac{{\bf D}_x^2(\tau_1.\tau_2)}{\tau_1\tau_2}+\frac{{\bf D}_x^2(\tau_3.\tau_4)}
{\tau_3\tau_4}\right]\left[\frac{{\bf SD}_x(\tau_1.\tau_1)}{\tau_1^2}-\frac{{\bf SD}_x(\tau_2.\tau_2)}{\tau_2^2}\right]\\
&-\frac{3}{2}\left[\frac{{\bf D}_x^2(\tau_1.\tau_1)}{\tau_1^2}-\frac{{\bf D}_x^2(\tau_2.\tau_2)}
{\tau_2^2}\right]\left[\frac{{\bf SD}_x(\tau_1.\tau_2)}{\tau_1\tau_2}-\frac{{\bf SD}_x(\tau_3.\tau_4)}{\tau_3\tau_4}\right]\\
&+\frac{3}{4}\left[\frac{{\bf D}_x^2(\tau_3.\tau_3)}{\tau_3^2}\frac{{\bf SD}_x(\tau_1.\tau_1)}
{\tau_1^2}-\frac{{\bf D}_x^2(\tau_1.\tau_1)}{\tau_1^2}\frac{{\bf SD}_x(\tau_3.\tau_3)}{\tau_3^2}\right]\\
&-\frac{3}{4}\left[\frac{{\bf D}_x^2(\tau_3.\tau_3)}{\tau_3^2}\frac{{\bf SD}_x(\tau_2.\tau_2)}
{\tau_2^2}-\frac{{\bf D}_x^2(\tau_2.\tau_2)}{\tau_2^2}\frac{{\bf SD}_x(\tau_3.\tau_3)}{\tau_3^2}\right]\\
&+\frac{3}{4}\left[\frac{{\bf D}_x^2(\tau_4.\tau_4)}{\tau_4^2}\frac{{\bf SD}_x(\tau_1.\tau_1)}
{\tau_1^2}-\frac{{\bf D}_x^2(\tau_1.\tau_1)}{\tau_1^2}\frac{{\bf SD}_x(\tau_4.\tau_4)}{\tau_4^2}\right]\\
&-\frac{3}{4}\left[\frac{{\bf D}_x^2(\tau_4.\tau_4)}{\tau_4^2}\frac{{\bf SD}_x(\tau_2.\tau_2)}{\tau_2^2}-\frac{{\bf D}_x^2(\tau_2.\tau_2)}
{\tau_2^2}\frac{{\bf SD}_x(\tau_4.\tau_4)}{\tau_4^2}\right]\\
&+\frac{3}{2}\left(\frac{{\bf D}_x^2(\tau_3.\tau_3)}{\tau_3^2}-\frac{{\bf D}_x^2(\tau_4.\tau_4)}
{\tau_4^2}\right)\left(\frac{\bar{{\bf D}}(\tau_3.\tau_1)}{\tau_3\tau_1}-\frac{\bar{{\bf D}}(\tau_3.\tau_2)}
{\tau_3\tau_2}-\frac{\bar{{\bf
D}}(\tau_4.\tau_1)}{\tau_4\tau_1}+\frac{\bar{{\bf
D}}(\tau_4.\tau_2)}{\tau_4\tau_2}\right) \, .
\label{12}
\end{split}
\ee
The second of equations in (\ref{8}) is obtained through the transformation
$\tau_1\leftrightarrow\tau_3$ and $\tau_2\leftrightarrow\tau_4$.

We will discuss a class of solutions of eqn. (\ref{12}) satisfying 
\be
\begin{split}
(4{\bf SD}_{t_3}-{\bf SD}_x^3)(\tau_a.\tau_a)&=0,\qquad {\rm for} \quad a=1,2,3,4\\
{\bf D}_x^2(\tau_1.\tau_2)&=0\\
{\bf D}_x^2(\tau_3.\tau_4)&=0\\
{\bf SD}_x(\tau_1.\tau_2)&=0\\
{\bf SD}_x(\tau_3.\tau_4)&=0\\
{\bf D}_x^2(\tau_a.\tau_a){\bf SD}_x(\tau_b.\tau_b)-{\bf D}_x^2(\tau_b.\tau_b){\bf SD}_x(\tau_a.\tau_a)&=0,
\qquad {\rm for} \quad a=3,4 \quad b=1,2\\
\bar{{\bf D}}(\tau_a.\tau_b)&=0, \qquad {\rm for} \quad a=3,4 \quad
b=1,2
\label{13}
\end{split}
\ee
Let all $\tau_i, \,i=1,...,4$ be of the form $1+ \Sigma$ where $\Sigma$ is a combination of exponential functions of
$\tilde \eta_a = 2k_a x +w_at + \zeta_a \theta, \,a=1,...,4$
with constant parameters $k_a, w_a$ and Grassmann parameters
$\zeta_a$.
In order to 
illustrate the method below we consider two explicit examples.
 
%%%%%%%%%%%%%%%%%%%%%%%%%%%%%%%%%%%%%%%%%%%%%%%%%%%%%%%%%%%%%%%%%%%%%%%%%%%%%%%%%%%%%%%%%%%%%%%%%%%%%%%%%%%%%%%%%%%%%%%%%%%%%%%%%%%%%%%%%%%%%%%%%%

\begin{itemize}
\item  {\it Two parameter solution}

Consider the following ansatz
\be\begin{split}
\tau_1&=1+\alpha_1 e^{\tilde{\eta}_1},\qquad \tau_2=1+\alpha_2 e^{\tilde{\eta}_2},\\
\tau_3&=1+\alpha_3 e^{\tilde{\eta}_3},\qquad
\tau_4=1+\alpha_4e^{\tilde{\eta}_4}\, .
\label{14}
\end{split}\ee

Using the relations,
\be\begin{split}
{\bf SD}_x^n(e^{\tilde{\eta}_1}.e^{\tilde{\eta}_2})&=(2k_1-2k_2)^n[-(\zeta_1-\zeta_2)+2\theta(k_1-k_2)]e^{\tilde{\eta}_1+\tilde{\eta}_2},\\
{\bf D}_x^n(e^{\tilde{\eta}_1}.e^{\tilde{\eta}_2})&=(2k_1-2k_2)^n
e^{\tilde{\eta}_1+\tilde{\eta}_2},\\
{\bf SD}_{t_3}^n(e^{\tilde{\eta}_1}.e^{\tilde{\eta}_2})&=(\omega_1-\omega_2)^n[-(\zeta_1-\zeta_2)+2\theta(k_1-k_2)]e^{\tilde{\eta}_1+\tilde{\eta}_2},\\
\bar{{\bf D}}(e^{\tilde{\eta}_1}.e^{\tilde{\eta}_2})&=2 (-\zeta_2k_1+\zeta_1 k_2)e^{\tilde{\eta}_1+\tilde{\eta}_2},
\label{15}
\end{split}\ee
we verify that eqns. (\ref{13})  are satisfied  if 
\be\begin{split}
k_2&=k_1, \qquad \qquad \,\,\,k_4=k_3,\\
\zeta_2&=\zeta_1, \qquad \qquad \,\,\,\,\zeta_4=\zeta_3,\\
\omega_1&=\omega_2={2k_1^3},\qquad \omega_3=\omega_4=2{k_3^3},\\
\alpha_2&=-\alpha_1, \qquad \quad
\,\,\,\,\alpha_4=-\alpha_3,\\
\zeta_3&=\frac{k_3}{k_1}\zeta_1.
\label{16}
\end{split}\ee
Explicitly, we find 
\be\begin{split}
u_1&=2k_1\alpha_1e^{\eta_1}\left(\frac{1}{1+\alpha_1e^{\eta_1}}+\frac{1}{1-\alpha_1e^{\eta_1}}\right),\\
u_3&=-2k_3\alpha_3e^{\eta_3}\left(\frac{1}{1+\alpha_3e^{\eta_3}}+\frac{1}{1-\alpha_3e^{\eta_3}}\right),\\
\psi_1&=-\zeta_1 \alpha_1e^{\eta_1}\left(\frac{1}{1+\alpha_1e^{\eta_1}}+\frac{1}{1-\alpha_1e^{\eta_1}}\right),\\
\psi_3&=-\zeta_1
\frac{k_3}{k_1}\alpha_3e^{\eta_3}\left(\frac{1}{1+\alpha_3e^{\eta_3}}+\frac{1}{1-\alpha_3e^{\eta_3}}\right),
\label{17}
\end{split}\ee
where $\zeta_1$ denotes the single Grassmann  parameter and 
\be
\eta_a=2(k_a x+{k_a^3}t_3), \qquad a=1,3.\label{18}
\ee

\item {\it Four parameter solution}

Consider the following  ansatz
\be\begin{split}
\tau_1&=1+\alpha_1 e^{\tilde{\eta}_1}+\alpha_2
e^{\tilde{\eta}_2}+\alpha_1\alpha_2
A_{1,2}e^{\tilde{\eta}_1+\tilde{\eta}_2},\\
\tau_2&=1+\beta_1 e^{\tilde{\eta}_1}+\beta_2
e^{\tilde{\eta}_2}+\beta_1\beta_2
B_{1,2}e^{\tilde{\eta}_1+\tilde{\eta}_2},\\
\tau_3&=1+\alpha_3 e^{\tilde{\eta}_3}+\alpha_4
e^{\tilde{\eta}_4}+\alpha_3\alpha_4
A_{3,4}e^{\tilde{\eta}_3+\tilde{\eta}_4},\\
\tau_4&=1+\beta_3 e^{\tilde{\eta}_3}+\beta_4
e^{\tilde{\eta}_4}+\beta_3\beta_4
B_{3,4}e^{\tilde{\eta}_3+\tilde{\eta}_4}\, .\label{19}
\end{split}\ee
Substituting in eqn.  (\ref{13}), we find
\be\begin{split}
\beta_s&=-\alpha_s, \quad \quad  \omega_s=2{k_s^3},\quad s=1,2,3,4
\label{20}
\end{split}\ee
\be\begin{split}
B_{i,j}&=A_{i,j}=\frac{(k_i-k_j)^2}{(k_i+k_j)^2}, \qquad
k_l\zeta_m=k_m\zeta_l,\label{21}
\end{split}\ee
for $(i=1,j=2)$, $(i=3,j=4)$ and $l,m=1,2,3,4$.
Conditions (\ref{16}) and (\ref{21}) justify the presence 
of  a single
Grassmann parameter $\zeta_1$.
In components we have
\be\begin{split}
\tau_k=\tau_k^a+\tau_k^b\zeta_1\theta,\qquad k=1,2,3,4\label{22}
\end{split}\ee
for which we obtain explicitly,
\be\begin{split}
\tau_1^a&=1+\alpha_1e^{\eta_1}+\alpha_2e^{\eta_2}+\alpha_1\alpha_2 A_{1,2}e^{\eta_1+\eta_2},\\
\tau_2^a&=1-\alpha_1e^{\eta_1}-\alpha_2e^{\eta_2}+\alpha_1\alpha_2 A_{1,2}e^{\eta_1+\eta_2},\\
\tau_3^a&=1+\alpha_3e^{\eta_3}+\alpha_4e^{\eta_4}+\alpha_3\alpha_4 A_{3,4}e^{\eta_3+\eta_4},\\
\tau_4^a&=1-\alpha_3e^{\eta_3}-\alpha_4e^{\eta_4}+\alpha_3\alpha_4 A_{3,4}e^{\eta_3+\eta_4},\\
\tau_1^b&=\frac{1}{k_1}\left(\alpha_1k_1e^{\eta_1}+\alpha_2k_2e^{\eta_2}+\alpha_1\alpha_2(k_1+k_2)A_{1,2}e^{\eta_1+\eta_2}\right),\\
\tau_2^b&=\frac{1}{k_1}\left(-\alpha_1k_1e^{\eta_1}-\alpha_2k_2e^{\eta_2}+\alpha_1\alpha_2(k_1+k_2)A_{1,2}e^{\eta_1+\eta_2}\right),\\
\tau_3^b&=\frac{1}{k_1}\left(\alpha_3k_3e^{\eta_3}+\alpha_4k_4e^{\eta_4}+\alpha_3\alpha_4(k_3+k_4)A_{3,4}e^{\eta_3+\eta_4}\right),\\
\tau_4^b&=\frac{1}{k_1}\left(-\alpha_3k_3e^{\eta_3}-\alpha_4k_4e^{\eta_4}+\alpha_3\alpha_4(k_3+k_4)A_{3,4}e^{\eta_3+\eta_4}\right),\label{24}
\end{split}\ee
where $\zeta_1$ is a constant fermionic parameter and
\be\begin{split}
\eta_a&=2(k_a x+ {k_a^3}t_3).
\label{25}
\end{split}\ee

\end{itemize}
 %%%%%%%%%%%%%%%%%%%%%%%%%%%%%%%%%%%%%%%%%%
 
 From (\ref{6}) and (\ref{10}) we find 
\be\begin{split}
u_1&=\partial_x\mathrm{ln}\left(\frac{\tau_1^a}{\tau_2^a}\right),
\qquad
\quad u_3=-\partial_x\mathrm{ln}\left(\frac{\tau_3^a}{\tau_4^a}\right),\\
\psi_1&= \zeta_1\left(\frac{\tau_2^b}{\tau_2^a}-\frac{\tau_1^b}{\tau_1^a}\right),
\qquad
\psi_3=\zeta_1\left(\frac{\tau_4^b}{\tau_4^a}-\frac{\tau_3^b}{\tau_3^a}\right),
\label{uu}
\end{split}\ee
 
Equation (\ref{uu}) together with equations 
(\ref{24})-(\ref{25})  provide a class of solutions 
for fields  $(u_1,\psi_1)$ and  $(u_3, \psi_3)$,  which  
satisfy both $N=1$ and $N=2$ super mKdV equations of motion 
since they also satisfy the non trivial relations like
\be \begin{split}
u_3^2 \pa_x \psi_1 &+ \h  \pa_x (u_3^2) \psi_1 + \pa_x(u_1 u_3) \psi_3 = 0, \\
u_1^2 \pa_x \psi_3 &+ \h  \pa_x (u_1^2) \psi_3 + \pa_x(u_1 u_3) \psi_1 = 0 \, .
\label{id}
\end{split}
\ee
 In general, identities (\ref{id}) follow directly from conditions (\ref{13}).
 
 We now discuss the corresponding soliton solutions for the $N=2$ super sinh-Gordon (\ref{2}) and (\ref{4}).
 Following the arguments of ref. \cite{nosso1}, where it
 was shown that the mKdV and sinh-Gordon models belong to the 
 same integrable hierarchy, and taking into account the 
 space-time dependence given in equation (\ref{i1}) 
 we relate solutions of both models to each other by 
 replacing in (\ref{25})
 \br
 k^3_a t_3 \rightarrow {k_a}^{-1} t_{-1}, \quad i.e., \quad \eta_a = 2(k_ax + k_a^{-1} t_{-1})
 \label{26}
 \er
 and $u_i = - \pa_x \phi_1, \;\; i=1,3$. Henceforth
 \be\begin{split}
\phi_1=-\mathrm{ln}\left(\frac{\tau_1^a}{\tau_2^a}\right), \quad \quad 
\phi_3=\mathrm{ln}\left(\frac{\tau_3^a}{\tau_4^a}\right). 
\label{gamma}
\end{split}\ee
 
 Plugging (\ref{24}) in (\ref{gamma}) and taking into account (\ref{26}) we
obtain fields $\phi_1, \phi_3$.  The fermionic fields $ \psi_1$ and $\psi_3$
are obtained from (\ref{uu}) with the space-time dependence given by (\ref{26}).
The auxiliary fields $\bar \psi_1$ and $\bar \psi_3$ are then solved by the
second of eqns. (\ref{2}) yielding,
 \be\begin{split}
\bar{\psi_1}&=\frac{1}{2}\left[\frac{(\partial_{t_{-1}}\psi_3)
\mathrm{sh}\phi_1\,\mathrm{sh}\phi_3-(\partial_{t_{-1}}\psi_1)
\mathrm{ch}\phi_1\,\mathrm{ch}\phi_3}{(\mathrm{ch}\phi_1\,\mathrm{ch}\phi_3)^2-(\mathrm{sh}\phi_1\,\mathrm{sh}\phi_3)^2}\right],\\
\bar{\psi_3}&=\frac{1}{2}\left[\frac{(\partial_{t_{-1}}\psi_3)
\mathrm{ch}\phi_1\,\mathrm{ch}\phi_3-(\partial_{t_{-1}}\psi_1)
\mathrm{sh}\phi_1\,\mathrm{sh}\phi_3}{(\mathrm{ch}\phi_1\,\mathrm{ch}\phi_3)^2-(\mathrm{sh}\phi_1\,\mathrm{sh}\phi_3)^2}\right]\,.
\label{27}
\end{split}\ee
It is interesting to consider as a particular  example, the case of $\a_1 =
\a_3 =0$ for which we obtain,  
\be\begin{split}
\bar{\psi}_1&=\frac{ 2 \zeta_1 e^{{{\eta }_2}}\,{{\alpha }_2}\,
    \left( 1 + e^{2\,{{\eta }_4}}\,
       {{{\alpha }_4}}^2 \right) }{{k_1}\,
    \left( -1 + e^{2\,{{\eta }_2}}\,
       {{{\alpha }_2}}^2 \right) \,
    \left( -1 + e^{2\,{{\eta }_4}}\,
       {{{\alpha }_4}}^2 \right) },\\
\bar{\psi}_3&=- \frac{ 2 \zeta_1 e^{{{\eta }_4}}\,
      \left( 1 + e^{2\,{{\eta }_2}}\,
         {{{\alpha }_2}}^2 \right) \,{{\alpha }_4}}
      {{k_1}\,\left( -1 +
        e^{2\,{{\eta }_2}}\,{{{\alpha }_2}}^2 \right) \,\left( -1 + e^{2\,{{\eta }_4}}\,
         {{{\alpha }_4}}^2 \right) } ,\label{ex1}
\end{split}\ee 
We have explicitly verified that the formulae (\ref{24}) with evolution parameter $t_{-1}$ given by
(\ref{26}) and general values of parameters $\a_i, i=1,2,3,4$ 
in equation (\ref{27}) indeed gives the solutions to the $N=2$ super 
sinh-Gordon  equations (\ref{2}).

We now relate the above solutions to solutions of 
the super $N=2$ KdV equation. 
Define  two spin $1/2$ superfields  $\Psi, i=1,2$ as
 \br
 \Psi_1 = \chi_1 + \chi_3, \quad \quad \Psi_2 = \chi_1 - \chi_3\, .
 \er
Eqn. (\ref{8}) gives
 \be \begin{split}
 4 \pa_{t_3} \Psi_1 &= D \left[ \pa_x^2 D \Psi_1 +3 \Psi_1 \pa_x \Psi_2
 D\Psi_2 - \h (D\Psi_1)^3 - \frac{3}{2} D\Psi_1 (D\Psi_2)^2\right], \\
 4 \pa_{t_3} \Psi_2 &= D \left[ \pa_x^2 D \Psi_2 +3 \Psi_2 \pa_x \Psi_1
 D\Psi_1 - \h (D\Psi_2)^3 - \frac{3}{2} D\Psi_2 (D\Psi_1)^2\right] \, .
 \end{split}\ee
 These equations, after time rescaling $t_3 \rightarrow -4t_3$, become eqns. (4.6) of ref. \cite{ik}.  Introducing the $N=2$ super Miura transformation given in eqn. (3.9) of \cite{ik}, i.e.
 \be
\begin{split}
 U &= D(\Psi_1+\Psi_2) - \Psi_1 \Psi_2 = 2 D\chi_1 +2 \chi_1 \chi_3, \\
 V &= \pa_x \Psi_2 -\Psi_2 D\Psi_1 = \pa_x \chi_1 -\pa_x \chi_3 -\chi_1 D\chi_1 - \chi_1 D \chi_3 + \chi_3 D \chi_1 + \chi_3 D \chi_3
 \label{uv}
 \end{split}
 \ee
 yields the $N=2$ super KdV equations of \cite{ik}, 
 \be
\begin{split}
 4\pa_{t_3} U &= \pa_x \left[ \pa_x^2 U +3 (DU) V - \h U^3\right], \\
 4\pa_{t_3} V &= \pa_x \left[ \pa_x^2 V -3 V(DV) + 3V\pa_x U - 
 \frac{3}{2}V U^2\right]
 \label{kdv}
\end{split}
\ee
for the spin 1 and $3/2$ superfields, respectively. 
Let the $U$ and $V$ be decomposed as follows
\[
U=U^b+\theta U^f \quad \quad V=V^f+\theta V^b\, ,
\]
with indices $b$ and $f$ referring to boson and fermion components,
respectively.  

Using (\ref{6})  in the r.h.s. of (\ref{uv}) we obtain
\be \begin{split}
U^b&=2(u_1+\psi_1\psi_3)  \\
U^f&=2(\partial_x\psi_1+u_3\psi_1+u_1\psi_3)\label{solution}
\end{split} \ee
and
\be \begin{split}
V^f&=\partial_x(\psi_1-\psi_3)-(u_1-u_3)(\psi_1-\psi_3) \\
V^b&=\partial_x(u_1+u_3)-(u_1^2-u_3^2)+(\psi_1-\psi_3)\partial_x(\psi_1+\psi_3)\,
. \label{ub}
\end{split}\ee
As an example we set $\a_2 = \a_4 =0$ in eqs. (\ref{24}).
Then equations (\ref{solution}) and (\ref{ub}) give 
\be\begin{split}
U^b&=\frac{8\,e^{{{\eta }_1}}\,{k_1}\,{{\alpha }_1}}
  {1 - e^{2\,{{\eta }_1}}\,{{{\alpha }_1}}^2},\\
U^f&=\frac{-8\,\zeta_1\,e^{{{\eta }_1}}\,{k_1}\,{{\alpha }_1}\,
    \left( 1 + e^{2\,{{\eta }_1}}\,
       {{{\alpha }_1}}^2 \right) }{{\left( -1 +
       e^{2\,{{\eta }_1}}\,{{{\alpha }_1}}^2 \right) }^
    2},\\
V^b&=\frac{-8\,\left( e^{{{\eta }_3}}\,{{k_3}}^2\,
       {\left( 1 + e^{{{\eta }_1}}\,{{\alpha }_1}
           \right) }^2 \,{{\alpha }_3} -
      e^{{{\eta }_1}}\,{{k_1}}^2\,{{\alpha }_1}\,
       {\left( 1 + e^{{{\eta }_3}}\,{{\alpha }_3}
           \right) }^2 \right) \,{}}{
    {\left( 1 + e^{{{\eta }_1}}\,{{\alpha }_1} \right)
        }^2\,{\left( 1 +
        e^{{{\eta }_3}}\,{{\alpha }_3} \right) }^2},\\
V^f&=\frac{4\,{{\zeta }_1}\,\left( e^{{{\eta }_3}}\,{{k_3}}^2\,
       {\left( 1 + e^{{{\eta }_1}}\,{{\alpha }_1}
           \right) }^2\,{{\alpha }_3} -
      e^{{{\eta }_1}}\,{{k_1}}^2\,{{\alpha }_1}\,
       {\left( 1 + e^{{{\eta }_3}}\,{{\alpha }_3}
           \right) }^2 \right) \,}{{k_1}\,
    {\left( 1 + e^{{{\eta }_1}}\,{{\alpha }_1} \right)
        }^2\,{\left( 1 +
        e^{{{\eta }_3}}\,{{\alpha }_3} \right) }^2}\, .
        \label{uuu}
\end{split}\ee

 Since we are considering the class of solutions with only one Grassmann
 constant parameter, the fermionic quadratic terms in (\ref{ub}) vanish
 identically. Furthermore, for the solutions given in (\ref{17}) and in
 (\ref{24}) since they satisfy $u_3 \psi_1 +u_1 \psi_3 =0$ (which can be
 checked in general using (\ref{13})), it follows that $U^b $ and $U^f$ depend
 only on $\alpha_1, \alpha_2$ and $V^b$ and $V^f$ depend upon $ \a_1, \a_2,
 \a_4$ and $\a_4$.  In particular, for $\a_3 = \a_4 = 0$ (with $t_3
 \rightarrow -4t_3, k \rightarrow k/2$), using algebraic computer methods, we
 have checked that our solutions agree with those found in \cite{gs}.  The
other solutions, like, for example those given in equation
(\ref{uuu}) with at least one of the parameters $\a_3, \a_4$ being 
different from zero are, as far as we are aware, new solutions.
 
It would be interesting to generalize this construction  to
involve multiple Grassmann constant parameters.  
In analogy to what was done for the corresponding $N=1$ hierarchy  in \cite{pla}, this may be accomplished 
in terms of vertex operators  and representations of the $\hat{sl}(2,2)$
affine algebra.

\vskip .4cm \noindent
{\bf Acknowledgements} \\
%\vskip .1cm \noindent
{  We are grateful to  CNPq and FAPESP  for 
financial support.}
The work is also supported by grant NSF PHY-0651694.

\end{document}